\documentclass[12pt]{article}
\usepackage{amsmath}
\usepackage{bm}
\usepackage[english]{babel}

\begin{document}
\begin{center}
\begin{large}
{\bf Effect of coordinate noncommutativity on the mass of a particle in a uniform field and the equivalence principle}\\
\end{large}
\end{center}

\centerline { Kh. P. Gnatenko \footnote{E-Mail address: khrystyna.gnatenko@gmail.com}, V. M. Tkachuk \footnote{E-Mail address: voltkachuk@gmail.com} }
\medskip
\centerline {\small \it Ivan Franko National University of Lviv, Department for Theoretical Physics,}
\centerline {\small \it 12 Drahomanov St., Lviv, 79005, Ukraine}

\begin{abstract}
We consider the motion of a particle in a uniform field in noncommutative space which is rotationally invariant. On the basis of exact calculations it is shown that there is an effect of coordinate noncommutativity on the mass of a particle. A particular case of motion of a particle in a uniform gravitational field is considered and the equivalence principle is studied. We propose the way to solve the problem of violation of the equivalence principle in the rotationally invariant noncommutative space.

\end{abstract}

\section{Introduction}

The idea of noncommutative structure of space was suggested by Heisenberg and later formalized by Snyder \cite{Snyder}. In recent years, due to the development of String Theory and Quantum Gravity  noncommutativity has received a considerable interest (see, for instance, \cite{Witten,Doplicher}).

In the canonical version of noncommutative space the coordinate and the momentum operators satisfy the following commutation relations
 \begin{eqnarray}
[X_{i},X_{j}]=i\hbar\theta_{ij},\label{form101}\\{}
[X_{i},P_{j}]=i\hbar\delta_{ij},\\{}
[P_{i},P_{j}]=0,\label{form10001}{}
\end{eqnarray}
with $\theta_{ij}$ being a constant antisymmetric matrix. Much attention has been devoted to studies of different physical problems in such a space, among them the hydrogen atom (see, for example, \cite{Chaichian,Ho,Chaichian1,Chair,Stern,Zaim2,Adorno,Khodja}), the Landau problem (see, for example, \cite{Nair,Bellucci,Dayi,Li,Dulat}), quantum mechanical system in a central potential \cite{Gamboa}, classical particle in a gravitational potential \cite{Romero,Mirza}, system of particles in a gravitational field \cite{Daszkiewicz}, motion of a body in a gravitational field and the equivalence principle \cite{Gnatenko1}, a particle in a gravitational quantum well in the case of canonical noncommutativity of coordinates and momentum-momentum noncommutativity \cite{Bertolami, Banerjee}, and many others.

It is worth noting that in the case of canonical version of noncommutative space (\ref{form101})-(\ref{form10001}) one faces the problem of rotational symmetry breaking \cite{Chaichian,Balachandran1}. Therefore, different classes of noncommutative algebras were considered to preserve the rotational symmetry (see, for instance, \cite{Gnatenko,Kupriyanov} and references therein).

It this paper we consider the motion of a particle in a uniform field in rotationally invariant noncommutative space which was proposed in our previous paper \cite{Gnatenko}. We show that there is an effect of noncommutativity on the mass of a particle. A particular case of motion of a particle in a uniform gravitational field is considered and the equivalence principle is studied. We propose the way to solve an important problem of violation of the equivalence principle in rotationally invariant noncommutative space and find the condition to recover this principle.

The paper is organized as follows. In Section \ref{rozd2}, we consider a noncommutative space with preserved rotational symmetry. In Section \ref{rozd3}, the motion of a particle in a uniform field in noncommutative space (\ref{form131})-(\ref{form13331}) is studied. It is shown that there is an effect of noncommutativity on the mass of a particle. In Section \ref{rozd4}, we consider the motion of a particle in a uniform gravitational field and study the equivalence principle.  Conclusions are presented in Section \ref{rozd5}.

\section{Noncommutative space with preserved rotational symmetry }\label{rozd2}

In previous paper \cite{Gnatenko} we studied the problem of rotational symmetry breaking in noncommutative space. To preserve this symmetry we considered a generalization of constant antisymmetric matrix $\theta_{ij}$ to a tensor constructed with the help of additional coordinates which are governed by a rotationally symmetric system. We supposed, for simplicity, that these coordinates are governed by the harmonic oscillator. According to the suggestion presented in \cite{Gnatenko} in the present paper we consider the tensor of noncommutativity which is constructed in the following form
\begin{eqnarray}
 \theta_{ij}=\frac{l_{0}}{\hbar}\varepsilon_{ijk} a_{k},  \label{form130}
 \end{eqnarray}
where $l_{0}$ is a constant with the dimension of length, $a_{i}$ are additional coordinates governed by the harmonic oscillator with parameters $m_{osc}$ and $\omega$
 \begin{eqnarray}
 H_{osc}=\frac{(p^{a})^{2}}{2m_{osc}}+\frac{m_{osc}\omega^{2} a^{2}}{2}.\label{form104}
 \end{eqnarray}
We consider the frequency of harmonic oscillator to be very large. In this case the distance between the energy levels of harmonic oscillator is very large too. Therefore, harmonic oscillator put into the ground state remains in it.

So, we consider the following rotationally invariant noncommutative algebra
\begin{eqnarray}
[X_{i},X_{j}]=i\varepsilon_{ijk} l_{0} a_{k},\label{form131}\\{}
[X_{i},P_{j}]=i\hbar\delta_{ij},\\{}
[P_{i},P_{j}]=0.\label{form13331}{}
\end{eqnarray}
 The coordinates $a_{i}$ and momenta $p^{a}_{i}$ satisfy the ordinary commutation relations $[a_{i},a_{j}]=0$, $[a_{i},p^{a}_{j}]=i\hbar\delta_{ij}$, $[p^{a}_{i},p^{a}_{j}]=0$. Also, $a_{i}$ commute with $X_{i}$ and $P_{i}$. Therefore, tensor of noncommutativity (\ref{form130}) commutes with $X_{i}$ and $P_{i}$ too. So, $X_{i}$, $P_{i}$ and $\theta_{ij}$ satisfy the same commutation relations as in the case of the canonical version of noncommutativity, moreover  algebra (\ref{form131})-(\ref{form13331}) is rotationally invariant.

 We would like to mention that additional
coordinates $a_{i}$ which form the tensor of noncommutativity can be treated as some
internal coordinates of a particle. Quantum fluctuations of these coordinates lead effectively
to a non-point-like particle, size of which is of the order of the Planck scale.

Coordinates $X_i$ and momenta $P_i$ can be represented by the coordinates $x_i$ and momenta $p_i$ which satisfy the ordinary commutation relations
\begin{eqnarray}
X_{i}=x_{i}-\frac{1}{2}\theta_{ij}{p}_{j},\label{form01010}\\
P_{i}=p_{i},\label{form01011}
\end{eqnarray}
where $\theta_{ij}$ is given by (\ref{form130}). Coordinates $x_{i}$ and momenta $p_{i}$ satisfy the following relations
\begin{eqnarray}
[x_{i},x_{j}]=0,\\{}
[p_{i},p_{j}]=0,\\{}
[x_{i},p_{j}]=i\hbar\delta_{ij},{}
\end{eqnarray}
and commute with $a_{i}$, $p^a_{i}$, namely  $[x_{i},a_{j}]=0$, $[x_{i},p^a_{j}]=0$, $[p_{i},a_{j}]=0$, $[p_{i},p^a_{j}]=0$. It is worth mentioning that coordinates $X_{i}$ do not commute with $p^a_{j}$. Taking into account (\ref{form130}) and (\ref{form01010}), we have  $[X_{i},p^a_{j}]=i\varepsilon_{ijk}l_{0}p_{k}/2$.

Explicit representation for coordinates $X_i$ (\ref{form01010}) and momenta $P_i$ (\ref{form01011}) guarantee that the Jacobi identity is satisfied. This can be easily checked for all possible triplets of operators.

Algebra (\ref{form131})-(\ref{form13331}) is manifestly rotationally invariant. It is clear that commutation relation (\ref{form131}) remains the same after rotation $X_{i}^{\prime}=U(\varphi)X_{i}U^{+}(\varphi)$, $a_{i}^{\prime}=U(\varphi)a_{i}U^{+}(\varphi)$. We have
\begin{eqnarray}
[X_{i}^{\prime},X_{j}^{\prime}]=i\varepsilon_{ijk}l_{0}a_{k}^{\prime},
\end{eqnarray}
 where the rotation operator reads $U(\varphi)=e^{\frac{i}{\hbar}\varphi({\bf n}\cdot{\bf\tilde{L}})}$. Here ${\bf\tilde{L}}$ is the total angular momentum which we define as follows
 \begin{eqnarray}
 {\bf\tilde{L}}=[{\bf x}\times{\bf p}]+[{\bf a}\times{\bf p}^{a}],
 \label{form500}
 \end{eqnarray}
 or taking into account (\ref{form130}), (\ref{form01010}) and (\ref{form01011}), we have
 \begin{eqnarray}
 {\bf\tilde{L}}=[{\bf X}\times{\bf P}]+\frac{l_0}{2\hbar}[{\bf P}\times[{\bf{a}}\times{\bf P}]]+[{\bf a}\times{\bf p}^{a}],
 \end{eqnarray}
 where ${\bf x}=(x_{1},x_{2},x_{3})$,  ${\bf X}=(X_{1},X_{2},X_{3})$ and ${\bf a}=(a_{1},a_{2},a_{3})$.
 It is worth mentioning that ${\bf\tilde{L}}$ satisfies commutation relations which are the same as in ordinary space, namely
$[X_{i},\tilde{L}_{j}]=i\hbar\varepsilon_{ijk}X_{k}$, $[P_{i},\tilde{L}_{j}]=i\hbar\varepsilon_{ijk}P_{k}$, $[a_{i},\tilde{L}_{j}]=i\hbar\varepsilon_{ijk}a_{k}$, $[p^{a}_{i},\tilde{L}_{j}]=i\hbar\varepsilon_{ijk}p^{a}_{k}$.

At the end of this Section we would like to note that taking into account (\ref{form01010}), we have that the operators $X_{i}$ depend on the momenta $p_i$ and therefore depend on the mass $m$. It is clear that operators $X_{i}$ do not depend on the mass in the case when $\theta_{ij}$ is proportional to $1/m$. So, in this case $X_{i}$ can be considered as a kinematic variables. This condition will be studied in details in Section \ref{rozd4}.

\section{Motion of a particle in a uniform field in noncommutative space with preserved rotational symmetry}\label{rozd3}

In this section we study the motion of a particle of mass $m$ in a uniform field in rotationally invariant noncommutative space (\ref{form131})-(\ref{form13331}). Let us consider the case when the field is pointed in the $X_{3}$ direction. So, the Hamiltonian of the particle is as follows
 \begin{eqnarray}
 H_{p}=\frac{P^{2}}{2m}+\kappa X_{3},
 \label{form888}
  \end{eqnarray}
where the uniform field is characterized by the factor $\kappa$. For example, in a particular case of motion of a charged  particle $q$ in the uniform electric field $E$ directed along the $X_3$ axis, factor $\kappa$ reads $\kappa=-qE$. In the case of motion of a particle of mass $m$ in the uniform gravitational field $g$ directed along the $X_3$ axis we have $\kappa=-mg$.

In rotationally invariant noncommutative space (\ref{form131})-(\ref{form13331}), because of definition of the tensor of noncommutativity (\ref{form130}), we have to take into account the
additional terms which correspond to the harmonic oscillator (\ref{form104}). So, we consider the total Hamiltonian as follows
 \begin{eqnarray}
H=H_{p}+H_{osc}=\frac{P^{2}}{2m}+\kappa X_{3}+\frac{(p^{a})^{2}}{2m_{osc}}+\frac{m_{osc}\omega^{2}a^{2}}{2}.\label{form1333}
\end{eqnarray}

Using representation (\ref{form01010}), (\ref{form01011}), we can rewrite Hamiltonian (\ref{form1333}) in the following form
\begin{eqnarray}
 H=\frac{p^{2}}{2m}+\kappa x_{3}+\frac{\kappa l_{0}}{2\hbar}\left(a_{1}p_{2}-a_{2}p_{1}\right) +\frac{(p^{a})^{2}}{2m_{osc}}+\frac{m_{osc}\omega^{2}a^{2}}{2}.\label{form600}
\end{eqnarray}
After algebraic transformations, we obtain
\begin{eqnarray}
 H=\left(1-\frac{\kappa^{2}l_{0}^{2}m}{4\hbar^{2}\omega^{2}m_{osc}}\right)\frac{p_{1}^{2}}{2m}+\left(1-\frac{\kappa^{2}l_{0}^{2}m}{4\hbar^{2}\omega^{2}m_{osc}}\right)\frac{p_{2}^{2}}{2m}+ \frac{p_{3}^{2}}{2m}+\kappa x_{3}+\frac{(p^{a})^{2}}{2m_{osc}}\nonumber\\+\frac{m_{osc}\omega^{2}}{2}\left(a_{1}+\frac{\kappa l_{0}}{2\hbar\omega^{2}m_{osc}}p_{2}\right)^{2}+\frac{m_{osc}\omega^{2}}{2}\left(a_{2}-\frac{\kappa l_{0}}{2\hbar\omega^{2}m_{osc}}p_{1}\right)^{2}+\frac{m_{osc}\omega^{2}a_{3}^{2}}{2}.\nonumber\\\label{form60110}
\end{eqnarray}
Hamiltonian (\ref{form60110}) can be rewritten as follows
\begin{eqnarray}
 H=\tilde{H}_{p}+\tilde{H}_{osc},\label{form601}
\end{eqnarray}
where
\begin{eqnarray}
\tilde{H}_{p}=\frac{p_{1}^{2}}{2m_{eff}}+\frac{p_{2}^{2}}{2m_{eff}}+ \frac{p_{3}^{2}}{2m}+\kappa x_{3}, \label{form61111}
\end{eqnarray}
with $m_{eff}$ being an effective mass which is defined as
\begin{eqnarray}
m_{eff}=m\left(1-\frac{\kappa^{2} l_{0}^{2}m}{4\hbar^{2}\omega^{2}m_{osc}}\right)^{-1},\label{form1}
\end{eqnarray}
and
\begin{eqnarray}
\tilde{H}_{osc}=\frac{(p^{a})^{2}}{2m_{osc}}+\frac{m_{osc}\omega^{2}q^{2}}{2}.\label{form2}
\end{eqnarray}
Here the components of ${\bf q}$ are the following
\begin{eqnarray}
q_{1}=a_{1}+\frac{\kappa l_{0}}{2\hbar\omega^{2}m_{osc}}p_{2},\\
q_{2}=a_{2}-\frac{\kappa l_{0}}{2\hbar\omega^{2}m_{osc}}p_{1},\\
q_{3}=a_{3}.
\end{eqnarray}
Note that $q_{i}$ satisfy the following commutation relations $[q_{i},q_{j}]=0$, $[q_{i},p^{a}_{j}]=i\hbar\delta_{ij}$, also $[q_{i},x_{j}]=-i\varepsilon_{ij3}\kappa l_{0}/(2m_{osc}\omega^{2})$, $[q_{i},p_{j}]=0$. So, Hamiltonian $\tilde{H}_{osc}$ corresponds to the tree-dimensional harmonic oscillator in the ordinary space.

It is important that $p_{1}^{2}/(2m_{eff})=\tilde{H}_{1}$, $p_{2}^{2}/(2m_{eff})=\tilde{H}_{2}$, $p_{3}^{2}/(2m)+\kappa x_{3}=\tilde{H}_{3}$ and $\tilde{H}_{osc}$ commute with each other.
The eigenfunctions of $H=\tilde{H}_{1}+\tilde{H}_{2}+\tilde{H}_{3}+\tilde{H}_{osc}$ (\ref{form601}) can be written as follows
 \begin{eqnarray}
\psi({\bf x},{\bf \tilde{q}})=C e^{ik_1x_1}e^{ik_2x_2}\psi^{(3)}(x_3)\psi^{\tilde{q}}({\bf \tilde{q}}),
\end{eqnarray}
 here $C$ is a constant, $k_1$ and $k_2$ are the components of the wave vector corresponding to the free motion of a particle in the perpendicular directions to the field direction,  $\psi^{(3)}(x_3)$ are well known eigenfunctions of $\tilde{H}_3$ which correspond to the motion of a particle in the field direction and can be written in terms of the Airy function, and $\psi^{\tilde{q}}({\bf \tilde{q}})$ are eigenfunctions of tree-dimensional harmonic oscillator with parameters $m_{osc}$ and $\omega$. The components of ${\bf\tilde{q}}$ read $\tilde{q}_1=a_{1}+\kappa l_{0}k_2/(2\omega^{2}m_{osc})$, $\tilde{q}_2=a_{2}-\kappa l_{0}k_1/(2\omega^{2}m_{osc})$ and $\tilde{q}_3=a_3$.

Let us write the eigenvalues of $H$ (\ref{form601}). Taking into account that the harmonic oscillator is in the ground state, we have
  \begin{eqnarray}
E=\frac{\hbar^2k_1^2}{2m_{eff}}+\frac{\hbar^2k_2^2}{2m_{eff}}+E_3+\frac{1}{2}\hbar\omega,\label{form20000}
\end{eqnarray}
where $E_3$ corresponds to the motion of a particle in the field direction.

Taking into account first two terms in (\ref{form20000}), we can conclude that there is an effect of coordinate noncommutativity on the mass of a particle. It is worth mentioning that noncommutativity has an effect on the motion of a particle in perpendicular directions to the direction of uniform field.  The motion of a particle in the field direction is governed by $\tilde{H}_{3}$  and is the same as in the ordinary space. Therefore, noncommutativity of coordinates causes the anisotropy of mass.

At the end of this section we would like to note that because of the rotationally
invariance obtained in this Section results can be easy generalized to the case of an arbitrary direction
of the uniform field.

In the next section, using obtained results, we consider a particular case of motion of a particle in a uniform gravitational field and study the equivalence principle.

\section{Motion of a particle in a uniform gravitational field. Equivalence principle}\label{rozd4}

In this section we study the free fall of a particle in the uniform gravitational field in noncommutative space (\ref{form131})-(\ref{form13331}). The gravitational field is directed along the $X_{3}$ axis and is characterized by the factor $g$. So, on the basis of the results obtained in Section 3, taking into account that $\kappa=-mg$, from (\ref{form601}), (\ref{form61111}) and (\ref{form1}) we have
\begin{eqnarray}
 H=\frac{p_{1}^{2}}{2m_{eff}}+\frac{p_{2}^{2}}{2m_{eff}}+\frac{p_{3}^{2}}{2m}-mg x_{3}+\tilde{H}_{osc},\label{form7}
\end{eqnarray}
with
\begin{eqnarray}
m_{eff}=m\left(1-\frac{l_{0}^{2}g^{2}m^3}{4\hbar^{2}\omega^{2}m_{osc}}\right)^{-1}.\label{form5}
\end{eqnarray}
 Taking into account  (\ref{form5}) and (\ref{form20000}), we can conclude that because of the term proportional to $m^{3}$ in  (\ref{form5}) the weak equivalence principle, also known as the universality of free
fall or the Galilean equivalence principle, is violated in noncommutative space (\ref{form131})-(\ref{form13331}).

In  previous paper \cite{Gnatenko1} we studied the problem of violation of the equivalence principle in two-dimensional noncommutative space in the case of canonical version of noncommutativity $[X_1,X_2]=i\hbar\theta$ with $\theta$ being a constant. We proposed the way to solve this problem. It was shown that the equivalence principle is recovered in the case when the following condition is satisfied
\begin{eqnarray}
\theta=\frac{\gamma}{m},\label{form03}
\end{eqnarray}
where $\theta$ is the parameter of noncommutativity, which corresponds to the particle of mass $m$, and $\gamma$ is a constant which takes the same value for all particles.

The problem of violation of the equivalence principle was also studied in deformed space with minimal length $[X,P]=i\hbar(1+\beta P^{2})$, where $\beta$ is the parameter of deformation \cite{Tkachuk}. It was shown that this problem is solved in the case when the parameter of deformation is completely determined by the mass of a particle in the following way $\beta m^2=const$.

Let us find the way to solve the problem of violation of the equivalence principle in rotationally invariant noncommutative space (\ref{form131})-(\ref{form13331}).
Taking into account (\ref{form20000}) and (\ref{form5}), it is clear that the necessary condition for the recovering of the equivalence principle is proportionality of the effective mass (\ref{form5}) to the mass of a particle. It can be realized when
\begin{eqnarray}
\frac{l_{0}^{2}m^3}{\omega^{2}m_{osc}}=A=const,\label{form00004}
\end{eqnarray}
where $A$ is a constant which takes the same value for particles of different masses. It is convenient to introduce the dimensionless constant
\begin{eqnarray}
\tilde{A}=A\frac{\omega_P^{2}}{l_{P}^{2}m_P^2},\label{form000042}
\end{eqnarray}
where $l_P$ and $m_P$ are the Planck length and the Planck mass, respectively, and $\omega_{P}$ is defined as $\hbar\omega_{P}=E_{P}$ with $E_{P}$ being the Planck energy.
Therefore, the effective mass reads
\begin{eqnarray}
m_{eff}=m\left(1-\frac{\tilde{A} l_{P}^{2}m_P^2 g^{2}}{4\hbar^{2}\omega_P^{2}}\right)^{-1}.\label{form51}
\end{eqnarray}

At the end of Section 2 we have noted that the operators $X_i$ depend on the mass of a particle. This also violates the equivalence principle. So, let us consider in details the condition which makes the operators $X_i$ independent on the mass of a particle (see the end of Section 2). For this purpose it is convenient to rewrite the tensor of noncommutativity in the following form
\begin{eqnarray}
 \theta_{ij}=\frac{l_{0}l_{osc}}{\hbar}\varepsilon_{ijk}\tilde{a}_{k}, \label{form120}
 \end{eqnarray}
where
 \begin{eqnarray}
 l_{osc}=\sqrt{\frac{\hbar}{m_{osc}\omega}},\label{form123}
  \end{eqnarray}
 and $\tilde{a}_{k}$ are the dimensionless coordinates $\tilde{a}_{k}=a_{k}/l_{osc}$.
 From (\ref{form01010}) and (\ref{form120}), it is clear that the operators $X_i$ do not depend on the mass of a particle in the case when
 \begin{eqnarray}
\frac{l_{0}l_{osc}}{l^2_P}=\tilde{\gamma}\frac{m_P}{m},\label{form501}
\end{eqnarray}
where $\tilde{\gamma}$ is a dimensionless constant which is the same for particles of different masses. For this constant we use notation $\tilde{\gamma}$ in order to distinguish it from $\gamma$ which was used in the canonical version of noncommutative space in \cite{Gnatenko1}.
Taking into account (\ref{form120}), it is clear that condition  (\ref{form501}) is similar to the condition (\ref{form03}) which gives the possibility to recover the equivalence principle in a two-dimensional space with canonical noncommutativity of coordinates.

So, conditions (\ref{form00004}) and (\ref{form501}) give the possibility to recover the equivalence principle in rotationally invariant noncommutative space (\ref{form131})-(\ref{form13331}).

It is worth mentioning that from  (\ref{form00004}), (\ref{form501}), using (\ref{form000042}) and (\ref{form123}), we obtain
\begin{eqnarray}
\omega=\frac{\tilde{\gamma}^2}{\tilde{A}}\omega_{P}\frac{m}{m_{P}}.\label{form514}
\end{eqnarray}
Therefore, both conditions are satisfied in the case when the frequency of harmonic oscillator is proportional to the mass of a particle.
Taking into account (\ref{form123}), (\ref{form501}) and (\ref{form514}), we also have that parameters $l_{osc}$, $l_0$ are inverse to $\sqrt{m}$, namely
\begin{eqnarray}
l_{osc}=l_P\sqrt{\frac{\tilde{A}m^2_P}{\tilde{\gamma}^2 m_{osc}m}},\label{form51400}\\
l_0=l_P\sqrt{\frac{\tilde{\gamma}^4m_{osc}}{\tilde{A}m}}.\label{form514000}
\end{eqnarray}

Let us estimate the value of $\tilde{A}$ and $\tilde{\gamma}$ constants. For this purpose we suppose that for the electron  $l_{0}=l_{P}$, $l_{osc}=l_{P}$, and $\omega=\omega_P$. In this case, taking into account (\ref{form123}), we have $m_{osc}=m_{P}$.
As a consequence, form (\ref{form00004}) and (\ref{form000042}), we find
 \begin{eqnarray}
\tilde{A}=\frac{m^3_e}{m^3_P}=7.3\times10^{-68},\label{form01}
 \end{eqnarray}
here $m_{e}$ is the mass of electron. Also, from (\ref{form501})  we obtain
 \begin{eqnarray}
\tilde{\gamma}=\frac{m_e}{m_P}=4.2\times10^{-23}.\label{form0100}
 \end{eqnarray}

At the end of this section we would like to mention that in the previous papers \cite{Gnatenko, Gnatenko3} the limit $\omega\rightarrow\infty$ was considered. Note that in this limit effect of noncommutativity on the mass of a particle tends to zero.

In this paper we consider a finite limit for $\omega$.
It is worth mentioning that fixing $\omega=\omega_P$, $l_{0}=l_{P}$, $l_{osc}=l_{P}$ for the electron,  from (\ref{form514}), (\ref{form51400}), (\ref{form514000})  we can calculate the parameters for a particle of mass $m_{i}$. We obtain $\omega^{(i)}=\omega_{P}m_{i}/m_{e}$, $l^{(i)}_{osc}=l_{P}\sqrt{m_{e}/m_{i}}$ and $l^{(i)}_{0}=l_{P}\sqrt{m_{e}/m_{i}}$.

 \section{Conclusion}\label{rozd5}

In this paper we have considered the motion of a particle in a uniform field in rotationally invariant noncommutative space (\ref{form131})-(\ref{form13331}).  All calculations presented here are exact. We have not made an approximation on the basis of general assumption that the parameter of noncommutativity is small, namely the parameter is of the order of squared Planck length. We have considered the suggestion presented in \cite{Gnatenko} to construct rotationally invariant noncommutative algebra by the generalization of a constant antisymmetric matrix $\theta_{ij}$ to a tensor defined by additional coordinates which are governed by harmonic oscillator.  It is worth mentioning that noncommutativity effects on the kinetic terms which correspond to the motion of a particle in perpendicular directions to the direction of the uniform field. The motion of a particle in these directions can be described with the help of effective mass. The motion of a particle in the field direction is the same as in the ordinary space. So, we have concluded that there is an effect of noncommutativity (\ref{form131}) on the mass of a particle in a uniform field and noncommutativity causes the anisotropy of mass.

The particular case of motion of a particle in the uniform gravitational field has been considered.
We have studied the problem of violation of the equivalence principle and found the conditions (\ref{form00004}), (\ref{form501}) to recover this principle in rotationally invariant noncommutative space (\ref{form131})-(\ref{form13331}). It is important to note that these conditions are in the agreement with the condition which gives the possibility to solve the problem of violation of the equivalence principle in a two-dimensional space with canonical noncommutativity of coordinates \cite{Gnatenko1}.

\section*{Acknowledgements}
The authors thank Dr. A. A. Rovenchak for a careful reading of the manuscript.


\begin{thebibliography}{0}
\bibitem{Snyder} H. Snyder, {\it Phys. Rev.} {\bf71}, 38 (1947).
\bibitem{Witten} N. Seiberg, E. Witten, {\it J. High Energy Phys.} {\bf9909}, 032 (1999).
\bibitem{Doplicher} S. Doplicher, K. Fredenhagen, J.E. Roberts, {\it Phys. Lett. B} {\bf331}, 39 (1994).
\bibitem{Chaichian} M. Chaichian, M.M. Sheikh-Jabbari, A. Tureanu, {\it Phys. Rev. Lett.}  {\bf86}, 2716 (2001).
\bibitem{Ho} Pei-Ming Ho, Hsien-Chung Kao, {\it Phys. Rev. Lett.} {\bf88}, 151602 (2002).
\bibitem{Chaichian1} M. Chaichian, M.M. Sheikh-Jabbari, A. Tureanu, {\it Eur. Phys. J. C} {\bf36}, 251 (2004).
\bibitem{Chair} N. Chair, M.A. Dalabeeh, {\it J. Phys. A, Math. Gen.} {\bf38}, 1553 (2005).
\bibitem{Stern} A. Stern, {\it Phys. Rev. Lett.} {\bf100}, 061601 (2008).
\bibitem{Zaim2} S. Zaim, L. Khodja, Y. Delenda, {\it Int. J. Mod. Phys. A}  {\bf26}, 4133 (2011).
\bibitem{Adorno} T.C. Adorno, M.C. Baldiotti, M. Chaichian, D.M. Gitman, A. Tureanu, {\it Phys. Lett.  B}  {\bf682}, 235 (2009).
\bibitem{Khodja} L. Khodja, S. Zaim, {\it Int. J. Mod. Phys. A} {\bf27}, 1250100 (2012).

\bibitem{Nair} V.P. Nair, A.P. Polychronakos, {\it Phys. Lett. B} {\bf505}, 267 (2001).
 \bibitem{Bellucci} S. Bellucci, A. Nersessian, C. Sochichiu, {\it Phys. Lett. B}  {\bf522}, 345 (2001).
\bibitem{Dayi} O.F. Dayi, L.T. Kelleyane, {\it Mod. Phys. Lett. A}  {\bf17}, 1937 (2002).
\bibitem{Li} Li Kang, Cao Xiao-Hua, Wang Dong-Yan, {\it Chin. Phys.}  {\bf15}, 2236 (2006).
\bibitem{Dulat} S. Dulat, K. Li, {\it Chin. Phys. C}  {\bf32}, 92 (2008).
\bibitem{Gamboa} J. Gamboa, M. Loewe, J.C. Rojas, {\it Phys. Rev. D} {\bf64}, 067901 (2001).
\bibitem{Romero} J.M. Romero, J.D. Vergara, {\it Mod. Phys. Lett. A} {\bf18}, 1673 (2003).
\bibitem{Mirza} B. Mirza, M. Dehghani, {\it Commun. Theor. Phys.}  {\bf42}, 183 (2004).
\bibitem{Daszkiewicz} M. Daszkiewicz, C.J. Walczyk, {\it Mod. Phys. Lett. A} {\bf26}, 819 (2011).
\bibitem{Gnatenko1}  Kh.P. Gnatenko, {\it Phys. Lett. A} {\bf377}, 3061 (2013).
\bibitem{Bertolami} O. Bertolami, J.G. Rosa, C.M.L. de Arag\~ao, P. Castorina, D. Zappal\`a, {\it Phys. Rev. D}  {\bf72}, 025010 (2005).
\bibitem{Banerjee} R. Banerjee, B.D. Roy, S. Samanta {\it Phys. Rev. D} {\bf74}, 045015 (2006).
\bibitem{Balachandran1} A.P. Balachandran, P. Padmanabhan, {\it J. High Energy Phys.}  {\bf1012}, 001 (2010).
\bibitem{Gnatenko} Kh.P. Gnatenko, V.M. Tkachuk, {\it Phys. Lett. A}  {\bf378}, 3509 (2014).
\bibitem{Kupriyanov} V.G. Kupriyanov, {\it Fortschr. Phys.} {\bf62}, 881 (2014).
\bibitem{Tkachuk} V.M. Tkachuk, {\it Phys. Rev. A} {\bf 86}, 062112 (2012).
\bibitem{Gnatenko3} Kh.P. Gnatenko, Yu.S. Krynytskyi, V.M. Tkachuk, {\it Mod. Phys. Lett. A}  {\bf30}, 1550033 (2015).


\end{thebibliography}
\end{document}